# Spatial Autocorrelation Equation Based on Moran's Index

Yanguang Chen

(Department of Geography, College of Urban and Environmental Sciences, Peking University, Beijing 100871, P.R. China. E-mail: chenyg@pku.edu.cn)

**Abstract**: Based on standardized vector and globally normalized weight matrix, Moran's index of spatial autocorrelation analysis has been expressed as a formula of quadratic form. Further, based on this formula, an inner product equation and outer product equation of the standardized vector can be constructed for Moran's index. However, the theoretical foundations and application direction of these equations are not yet clear. This paper is devoted to exploring the inner and outer product equations of Moran's index. The methods include mathematical derivation and empirical analysis. The results are as follows. First, based on the inner product equation, two spatial autocorrelation models can be constructed. One bears constant terms, and the other bear no constant term. The spatial autocorrelation models can be employed to calculate Moran's index by regression analysis. Second, the inner and outer product equations can be used to improve Moran's scatterplot. The normalized Moran's scatterplot can show more geospatial information than the conventional Moran's scatterplot. A conclusion can be reached that the spatial autocorrelation models are useful spatial analysis tools, complementing the uses of spatial autocorrelation coefficient and spatial autoregressive models. These models are helpful for understanding the boundary values of Moran's index and spatial autoregressive modeling process.

# 1. Introduction

Spatial autocorrelation measures can be expressed as simple forms based on standardize vector and normalized weight matrix. Then the basic measures such as Moran's index can be integrated into a pair of equations: one is inner product equation, and other is outer product equation. This two equations can be used to estimate Moran's index and normalize Moran's scatterplot. The scatterplot was proposed by Anselin (1996) for local spatial autocorrelation analysis. It is easy to understand and make use of the outer product equation, in which the nonzero eigenvalue is Moran's index. However, the inner product equation is difficult to understand in a simple way. Formally, the eigenvalue in the inner product equation is equal to Moran's index. However, empirically, this characteristic parameter is not a numerical value, but a set of eigenvalues. Moran's index comes between the maximum eigenvalue and the minimum eigenvalue. On the other hand, by means of linear regressive analysis, the characteristic parameter becomes the regressive coefficient and gives the value of Moran's index (Chen, 2013).

In fact, the inner product equation of Moran's index deserves further research. Based on the inner product equation, two spatial autocorrelation models can be constructed. One is the linear model with a constant term, and the other is the linear model without a constant term. Derivation of the formulae of models' parameters is helpful for developing theory and method of spatial autocorrelation analysis. First, the results can be used to help us understand Moran's index deeply and improve normalized Moran's scatterplot. Second, the results can be used to analyze the boundary values of Moran's index from different angles of view. Third, the results are useful for understanding the related spatial autoregressive modeling from the perspective of spatial autocorrelation. This paper is devoted to clarifying the mathematical structure of parameters of spatial autocorrelation models based on Moran's index. The rest parts are organized as follows. In Section 2, two sets of parameter expressions are derived from the spatial autocorrelation models by using four methods. In Section 3, an empirical analysis is made to verify the results of theoretical derivation. In Section 4, several questions are discussed, and finally, in Section 5, the discussion is concluded by summarizing the main points of this study.

## 2. Theoretical results

### 2.1 Preparatory equations

Starting from the concise mathematical form, we can derive new clear relations for spatial autocorrelation analysis. Suppose that there are $n$ elements such as cities in a geographical region. We can use Moran's index to reflect the extent of spatial autocorrelation of the $n$ geographical elements (Haggett *et al*, 1977; Moran, 1950; Odland, 1988). Based on population standardized size variable and globally normalized spatial weight matrix, Moran's index can be express as (Chen, 2013)

$$I = \mathbf{z}^T \mathbf{W} \mathbf{z}, \qquad (1)$$

where *I* refers to Moran's index (Moran's *I* for short), $\mathbf{z}=[z_1, z_2,…, z_n]^T$ is the standardized size vector by *z*-score, $\mathbf{W}=[w_{ij}]$ is a $n \times n$ globally normalized symmetric spatial weight matrix, and the superscript T implies transposition of matrix or vector (*i*, *j*=1,2,…,*n*). Equation (1) is a kind of quadratic form in mathematics. The main properties of globally normalized spatial weight matrix are as follows: (1) global normalization, that is, the sum of element values in **W** is 1; (2) symmetry, that is, $\mathbf{W}^T=\mathbf{W}$; (3) Non-negativity, that is, all the values in the matrix are greater than or equal to 0. The properties of standardized variable are as below: the mean of **z** is 0, and standard deviation of **z** is 1. Moreover, we have the following relation

$$\mathbf{z}^T \mathbf{z} = \mathbf{o}^T \mathbf{o} = n, \qquad (2)$$

where $\mathbf{o}=[1, 1,…,1]^T$ is a one vector consisting of *n* ones. It is easy to prove or verify equation (2) by using the knowledge of linear algebra.

Base on Moran's index, a set of spatial autocorrelation equations can be constructed. The first equation is based on the outer product of **z**. Multiplying left equation (1) by **z** yields

$$\mathbf{z}\mathbf{z}^T \mathbf{W} \mathbf{z} = I\mathbf{z}, \qquad (3)$$

which is a characteristic equation of Moran's index. This suggests that the absolute value of *I* is the only nonzero eigenvalue of the spatial correlation matrix $\mathbf{z}\mathbf{z}^T\mathbf{W}$, and **z** is the corresponding eigenvector. The second equation is based on the inner product of **z**. An approximate characteristic function can be expressed as follows

$$\mathbf{z}^T \mathbf{z} \mathbf{W} \mathbf{z} = n\mathbf{W}\mathbf{z} = I\mathbf{z}, \qquad (4)$$

which is a pure theoretical relation. In theory, *I* is the eigenvalue of rescaled spatial weight matrix $\mathbf{z}^T\mathbf{z}\mathbf{W}$, but in empirical analysis, the eigenvalue is a set of values and can be expressed as $\lambda_i$ (*i*=1, 2,…, *n*). It can be proved that the value of *I* comes between the minimum eigenvalue $\lambda_{min}$ and the maximum eigenvalue $\lambda_{max}$ of $\mathbf{z}^T\mathbf{z}\mathbf{W}$, that is, $\lambda_{min} \leq I \leq \lambda_{max}$. By using the least squares calculation, we can find the following linear equation:

$$\mathbf{z}^T \mathbf{z} \mathbf{W} \mathbf{z} = n\mathbf{W}\mathbf{z} = (\mathbf{o}^T\mathbf{W}\mathbf{z})\mathbf{o} + I\mathbf{z}, \qquad (5)$$

where $\mathbf{o}^T\mathbf{W}\mathbf{z}=(\mathbf{W}\mathbf{z})^T\mathbf{o}$ is a constant. Equations (4) and (5) are in fact inverse functions of spatial autoregressive models and can be regarded as inverse spatial autoregressive functions.

The above mathematical equations are helpful to improve spatial autocorrelation analysis and prepare for understanding spatial autoregressive modeling. Equations (3) and (4) can be used to generate a normalized Moran's scatterplot (Chen, 2013). Equation (4) gives scattered points, equation (3) gives a theoretical trend line, and both equation (4) and equation (5) give an empirical trend lines. The slopes of the trend lines gives the Moran's index value. However, the theoretical foundation of equations (4) and (5) is not yet clear. The outer product equation of Moran's index is easy to understand, while the corresponding inner product equation is relatively complex. For observational data from the real world, equation (5) should be replaced by an empirical relation as below

$$n\mathbf{Wz} = a\mathbf{o} + b\mathbf{z} + \mathbf{e}, \qquad (6)$$

where $a$ refers to intercept, i.e., the constant term of the regression equation, and $b$ denotes the slope, i.e., the regressive coefficient in a narrow sense, $\mathbf{e}$ is a residuals set, representing an error term. A simple spatial autoregressive model can be obtained by exchanging the positions of independent variables and dependent variables in equation (6). Now, it is necessary to prove the follow relations: $a = \mathbf{o}^T\mathbf{Wz} = (\mathbf{Wz})^T\mathbf{o}$, $b=I$, which are related to the theoretical essence of the Moran's autocorrelation equations.

## 2.2 Regressive coefficients of autocorrelation models

Using the principle of linear algebra and related knowledge of linear regressive analysis, we can derive the parameter expressions for the inner product equations based on Moran's index. Equation (6) left multiplied by the transpose of vector $\mathbf{z}$ and $\mathbf{Wz}$, respectively, yields a pair of equations as below

$$n\mathbf{z}^T\mathbf{Wz} = a\mathbf{z}^T\mathbf{o} + b\mathbf{z}^T\mathbf{z} + \mathbf{z}^T\mathbf{e}, \qquad (7)$$

$$n(\mathbf{Wz})^T\mathbf{Wz} = a(\mathbf{Wz})^T\mathbf{o} + b(\mathbf{Wz})^T\mathbf{z} + (\mathbf{Wz})^T\mathbf{e}. \qquad (8)$$

The sum of the values of a normalized vector equals 0, that is, $\mathbf{z}^T\mathbf{o}=0$. On the other hand, the independent variables $z$ and residuals $e$ are orthogonal to each other, that is

$$\mathbf{z}^T\mathbf{e} = \mathbf{z}^T\mathbf{o} = 0. \qquad (9)$$

The inner product between $Wz$ and residuals $e$ can be expressed as

$$\gamma = (\mathbf{Wz})^T\mathbf{e}, \qquad (10)$$

which $\gamma$ is a parameter and will be further explained later. According to equations (1), (2) and (10),

equations (7) and (8) can be changed to

$$nI = 0a + nb,  \quad (11)$$

$$n(\mathbf{Wz})^T \mathbf{Wz} - \gamma = (\mathbf{Wz})^T \mathbf{o}a + Ib.  \quad (12)$$

From equation (11), the slope can be derived as

$$b = I.  \quad (13)$$

This proves that the slope is equal to Moran's index. From equation (12) we derive the preliminary expression of the intercept as follows

$$a = \frac{n(\mathbf{Wz})^T \mathbf{Wz} - I^2 - \gamma}{(\mathbf{Wz})^T \mathbf{o}},  \quad (14)$$

in which $\gamma$ will be proved to be the variance of residuals later.

## 2.3 Intercept and slope

It can be further proved that the value of the regression coefficient representing the slope remains unchanged regardless of whether the intercept is retained in the model. Concretely speaking, if the constant term in equation (5) is removed or set as 0, we have the form of equation (4). Equation (4) and equation (5) share the same slope. First of all, we examine the model with constant term, equation (6). Letting $x=z$ and $y=nWz$, we can apply the formulae of coefficients for one variate linear regressive models to equation (6) (Appendix 1). Considering that the mean of $z_i$ is zero, we have

$$b = \frac{\sum_{i=1}^{n} z_i (n\mathbf{Wz})_i - \langle n\mathbf{Wz} \rangle \sum_{i=1}^{n} z_i}{\sum_{i=1}^{n} z_i^2} = \frac{n \sum_{i=1}^{n} z_i (\mathbf{Wz})_i}{\sum_{i=1}^{n} z_i^2} = \mathbf{z}^T \mathbf{Wz} = I,  \quad (15)$$

where ‹•› represents averaging "•", $i=1,2,3,\ldots,n$. In equation (15), $\sum z_i = \mathbf{z}^T \mathbf{o} = 0$, $\sum z_i^2 = \mathbf{z}^T \mathbf{z} = n$. This proves that the slope is just Moran's index from a new perspective. The constant term can be calculated by the means of $z$ and $nWz$. Thus we have

$$a = \frac{1}{n} \sum_{i=1}^{n} (n\mathbf{Wz})_i - b \frac{1}{n} \sum_{i=1}^{n} z_i = \sum_{i=1}^{n} (\mathbf{Wz})_i = (\mathbf{Wz})^T \mathbf{o},  \quad (16)$$

which indicates that the intercept is just the mean of $nWz$. It can be proved that if Moran's index $I=b=1$, we have $a=(\mathbf{Wz})^T\mathbf{o}=0$. This suggests the intercept $a$ can also reflect spatial autocorrelation. The stronger the autocorrelation, the closer the intercept approaches 0.

Next, let's examine the model without constant term. This model can be given by revising equation (4). The expression is as follows

$$\mathbf{z}^T\mathbf{z}\mathbf{W}\mathbf{z} = n\mathbf{W}\mathbf{z} = b\mathbf{z} + \mathbf{e}^*, \tag{17}$$

where $\mathbf{e}^*$ represents the residuals term in the zero-intercept model. In this instance, the slope can be given by the ratio of the inner product between $z$ and $nWz$ to the inner product of $z$, that is

$$b = \frac{\sum_{i=1}^{n} z_i (n\mathbf{W}\mathbf{z})_i}{\sum_{i=1}^{n} z_i^2} = \frac{n\mathbf{z}^T\mathbf{W}\mathbf{z}}{\mathbf{z}^T\mathbf{z}} = \mathbf{z}^T\mathbf{W}\mathbf{z} = I. \tag{18}$$

Comparing equation (18) with equation (15) shows that the slope of the linear model without intercept is the same as that of the linear model with intercept. This suggests that the existence of intercept does not influence the value of slope, and thus the intercept and slope can be used to describe spatial autocorrelation independently. Especially, the regressive coefficient can be connected to the eigenvalue of $n\mathbf{W}$.

For the model with constant term, we have two different expressions of the intercept. One is equation (14), and the other is equation (18). Combining equation (14) and equation (18) yields

$$a = \frac{n(\mathbf{W}\mathbf{z})^T\mathbf{W}\mathbf{z} - I^2 - \gamma}{(\mathbf{W}\mathbf{z})^T\mathbf{o}} = (\mathbf{W}\mathbf{z})^T\mathbf{o}. \tag{19}$$

From equation (19) it follows

$$\gamma = n(\mathbf{W}\mathbf{z})^T\mathbf{W}\mathbf{z} - ((\mathbf{W}\mathbf{z})^T\mathbf{o})^2 - I^2. \tag{20}$$

On the other hand, the inner product of the residuals vector is

$$\mathbf{e}^T\mathbf{e} = (n\mathbf{W}\mathbf{z} - a\mathbf{o} - b\mathbf{z})^T(n\mathbf{W}\mathbf{z} - a\mathbf{o} - b\mathbf{z}). \tag{21}$$

Expanding and rearranging equation (21) yields (Appendix 2)

$$\mathbf{e}^T\mathbf{e} = n(n(\mathbf{W}\mathbf{z})^T\mathbf{W}\mathbf{z} - ((\mathbf{W}\mathbf{z})^T\mathbf{o})^2 - I^2). \tag{22}$$

The variance of the residuals is obtained by dividing the inner product of $e$ by $n$. Comparing equation (22) with equation (20), we have

$$\gamma = \frac{1}{n}\mathbf{e}^T\mathbf{e} = n(\mathbf{W}\mathbf{z})^T\mathbf{W}\mathbf{z} - ((\mathbf{W}\mathbf{z})^T\mathbf{o})^2 - I^2 = (\mathbf{W}\mathbf{z})^T\mathbf{e} = \sigma_e^2. \tag{23}$$

where $\sigma_e$ denotes the population standard deviation of the residuals $e$. Substituting equation (23) into equation (20) yields

$$n(\mathbf{W}\mathbf{z})^T\mathbf{W}\mathbf{z} - I^2 - \sigma_e^2 = ((\mathbf{W}\mathbf{z})^T\mathbf{o})^2. \tag{24}$$

This is a useful relation for understanding the boundary values of Moran's index.

## 2.4 Least squares algorithm

The premise of application of a mathematical model to real problems lies in effective algorithm. The principal algorithm of ordinary linear regressive analysis is the least squares method. The above reasoning has involved the least squares method. It can be demonstrated that the least squares principle can be applied to estimation of the values of Moran's index and related parameters. Based on matrix transformation, the regressive coefficients of equation (6) can be computed by

$$\mathbf{B} = \left[\begin{bmatrix} \mathbf{o}^T \\ \mathbf{z}^T \end{bmatrix} \begin{bmatrix} \mathbf{o} & \mathbf{z} \end{bmatrix}\right]^{-1} \begin{bmatrix} \mathbf{o}^T \\ \mathbf{z}^T \end{bmatrix} n\mathbf{Wz} = \begin{bmatrix} \mathbf{o}^T\mathbf{o} & \mathbf{o}^T\mathbf{z} \\ \mathbf{z}^T\mathbf{o} & \mathbf{z}^T\mathbf{z} \end{bmatrix}^{-1} \begin{bmatrix} n\mathbf{o}^T\mathbf{Wz} \\ n\mathbf{z}^T\mathbf{Wz} \end{bmatrix}, \tag{25}$$

where $\mathbf{B}$ refers to the coefficient vector. As indicated above, $\mathbf{o}^T\mathbf{z}=0$, $\mathbf{o}^T\mathbf{o}=n$, $\mathbf{z}^T\mathbf{Wz}=I$. According to the principle of linear regression analysis, equation (25) can be changed to

$$\mathbf{B} = \begin{bmatrix} n & 0 \\ 0 & n \end{bmatrix}^{-1} \begin{bmatrix} n\mathbf{o}^T\mathbf{Wz} \\ n\mathbf{z}^T\mathbf{Wz} \end{bmatrix} = \frac{1}{n^2}\begin{bmatrix} n & 0 \\ 0 & n \end{bmatrix}\begin{bmatrix} n\mathbf{o}^T\mathbf{Wz} \\ n\mathbf{z}^T\mathbf{Wz} \end{bmatrix} = \begin{bmatrix} \mathbf{o}^T\mathbf{Wz} \\ I \end{bmatrix}. \tag{26}$$

Apparently, the vector $\mathbf{B}$ gives two regression parameters: one is the constant term, and the other is the autocorrelation coefficients. Equation (26) corresponds to equations (15) and (16).

The above process can also be processed by means of determinant operation. Based on equations (11) and (12), three determinants can be constructed as follows

$$A = \begin{vmatrix} nI & n \\ n(\mathbf{Wz})^T\mathbf{Wz}-\gamma & I \end{vmatrix}, \quad B = \begin{vmatrix} 0 & nI \\ (\mathbf{Wz})^T\mathbf{o} & n(\mathbf{Wz})^T\mathbf{Wz}-\gamma \end{vmatrix}, \quad C = \begin{vmatrix} 0 & n \\ (\mathbf{Wz})^T\mathbf{o} & I \end{vmatrix}.$$

According to Cramer's rule in linear algebra, the regressive coefficients can be calculated by the following formulae:

$$a = \frac{A}{C} = \frac{n(\mathbf{Wz})^T\mathbf{Wz} - I^2 - \sigma_e^2}{(\mathbf{Wz})^T\mathbf{o}}, \tag{27}$$

$$b = \frac{B}{C} = \frac{-nI(\mathbf{Wz})^T\mathbf{o}}{-n(\mathbf{Wz})^T\mathbf{o}} = I, \tag{28}$$

which are equivalent to the results from matrix transformation. Equations (27) and (28) corresponds to equations (13) and (14). Due to $\mathbf{z}^T\mathbf{e}=0$, determinant operation is a little indirect and superfluous where practical application is concerned. However, in theory, combining the matrix transformation results and determinant operation results, we can find another way to derive equation (24), which provides another way of understanding the bounds of Moran's index.

## 2.5 Parameter bounds

The lower and upper boundary values of Moran's index is a controversial issue. By analogy with the bounds of Pearson correlation coefficient and autocorrelation coefficients of time series, the bounds of Moran's index was regarded as coming between -1 and 1. However, another findings is that Moran's index varies from the minimum eigenvalue to the maximum eigenvalue of the $n$ times of globally normalized spatial weight matrix (de Jong *et al*, 1984; Tiefelsdorf and Boots, 1995; Xu, 2021). This paper does not give the final judgment of the boundary values of Moran's index, but constructs the criterion from three different angles of view. First, *the angle of spatial weight matrix*. In light of equations (1) and (2), we have

$$I = \frac{\mathbf{z}^T(n\mathbf{W})\mathbf{z}}{\mathbf{z}^T\mathbf{z}} = \frac{\mathbf{z}^T(\mathbf{z}^T\mathbf{z}\mathbf{W})\mathbf{z}}{\mathbf{z}^T\mathbf{z}}. \tag{29}$$

According to the principle of Rayleigh quotient (Xu, 2021), we get the first value range of Moran's index as follows

$$\lambda_{\min} \leq I \leq \lambda_{\max}, \tag{30}$$

in which $\lambda$ denotes the eigenvalues of $n\mathbf{W}$, $\lambda_{\max}$ and $\lambda_{\min}$ are the maximum and minimum eigenvalues of $n\mathbf{W}$, respectively. More than one method results in the inference that the value of Moran's index is determined by the minimum and maximum eigenvalues of spatial weight matrix (de Jong *et al*, 1984; Tiefelsdorf and Boots, 1995).

Second, *the angle of the inner product of spatial weight matrix*. The inner product of spatial weight matrix is $\mathbf{W}^T\mathbf{W}$. From the inner product equation, equations (5), it follows

$$I^2 + ((\mathbf{W}\mathbf{z})^T\mathbf{o})^2 = \frac{\mathbf{z}^T(n^2\mathbf{W}^T\mathbf{W})\mathbf{z}}{\mathbf{z}^T\mathbf{z}} = \frac{\mathbf{z}^T(n\mathbf{W})^T(n\mathbf{W})\mathbf{z}}{\mathbf{z}^T\mathbf{z}}. \tag{31}$$

So we have the second value range of parameters as follows

$$\lambda^*_{\min} \leq I^2 + ((\mathbf{W}\mathbf{z})^T\mathbf{o})^2 \leq \lambda^*_{\max}, \tag{32}$$

where $\lambda^*$ refers to the eigenvalues of $(n\mathbf{W})^T(n\mathbf{W})$, $\lambda^*_{\max}$ and $\lambda^*_{\min}$ are the maximum and minimum eigenvalues of $(n\mathbf{W})^T(n\mathbf{W})$, respectively. This represents a pure theoretical expression. In the mathematical world, the error term can be removed, and thus $\sigma_e=0$. However, in the real world, the errors cannot be ignored. Thus, based on equation (24), equation (32) can be revised as

$$\lambda_{\min}^{*} \leq I^2 + ((\mathbf{Wz})^T \mathbf{o})^2 + \sigma_e^2 \leq \lambda_{\max}^{*}, \tag{33}$$

which represents an empirical expression for the second boundary values.

Third, *the angle of the outer product of weighted size vector*. The weighted vector is $\mathbf{Wz}$, and the corresponding outer product is $\mathbf{Wz}(\mathbf{Wz})^T = \mathbf{Wzz}^T\mathbf{W}$. From the outer product equation, equation (3), we can derive the following relation

$$(\mathbf{Wz})^T \mathbf{zz}^T \mathbf{Wz} = I(\mathbf{Wz})^T \mathbf{z} = I^2, \tag{34}$$

which apparently can be obtained from equation (1) by taking square. Considering equation (2), we have

$$I^2 = \frac{\mathbf{z}^T(n\mathbf{W}^T\mathbf{zz}^T\mathbf{W})\mathbf{z}}{\mathbf{z}^T\mathbf{z}} = \frac{\mathbf{z}^T\mathbf{W}^T\mathbf{z}^T\mathbf{zzz}^T\mathbf{Wz}}{\mathbf{z}^T\mathbf{z}}. \tag{35}$$

Then we have the third set of value range of Moran's index as follows

$$\lambda_{\min}^{**} \leq I^2 \leq \lambda_{\max}^{**}, \tag{36}$$

where $\lambda^{**}$ denotes the eigenvalues of $n\mathbf{W}^T\mathbf{zz}^T\mathbf{W}$, $\lambda^{**}_{\max}$ and $\lambda^{**}_{\min}$ represent the maximum and minimum eigenvalues of $n\mathbf{W}^T\mathbf{zz}^T\mathbf{W}$, respectively. It can be proved that $\lambda^{**}_{\min} = 0$, $\lambda^{**}_{\max} = n(\mathbf{Wz})^T\mathbf{Wz}$, which suggests that $I^2 \leq n(\mathbf{Wz})^T\mathbf{Wz}$.

## 3. Empirical analysis

### 3.1 Study area and data

The main aim of this work is at exploring the theoretical foundation of spatial autocorrelation models based on Moran's index. Nevertheless, the results of mathematical derivation need to be testified by observational data from the real world. If and only if a reasoning result is consistent with the calculated results based on observed data, it is really valid. As we know, the success of natural sciences lies in their great emphasis on the role played by quantifiable data and their interplay with models (Louf and Barthelemy, 2014). So do social sciences to a great degree today. To evaluate the theoretical reasoning results, several sets of observational data will be utilized to verify the models and relations given in Section 2. The study area is Beijing-Tianjin-Hebei (BTH) region of China, including Beijing Municipality, Tianjin Municipality, and Hebei Province. It is sometimes termed Jing-Jin-Ji (JJJ) region in literature. The study region includes 13 cities at and above the prefecture level and 22 county-level cities. That is to say, there are 35 cities in total in our study area ($n$=35). Three sources of observational data are available. The spatial distances are measured by traffic mileage map, and this dataset was extracted by ArcGIS. City sizes were measured by two indicators. One is urban census population in 2000 (the fifth census) and 2010 (the sixth census), and the other is urban nighttime light (NTL) intensity (Table 1). The data of NTL intensity can be reflected by the total number of NTLs within built-up area of cities in the BTH region (Chen and Long, 2021; Long and Chen, 2019). The spatial proximity is defined by the inverse function of distance, i.e., $v_{ij}$=1/$r_{ij}$, where $r_{ij}$ denotes the traffic mileage between city $i$ and city $j$. Therefore, the spatial contiguity matrix can be expressed as **V**=[$v_{ij}$]=[1/$r_{ij}$], in which the diagonal elements are defined as zero. That is to say, if $i$=$j$, then $v_{ij}$ =0, or else, $v_{ij}$=1/$r_{ij}$. Globally normalizing **V** results in a spatial weight matrix **W**=**V**/**V**$_0$=[$w_{ij}$], where $w_{ij}$ = $v_{ij}$/ **V**$_0$, and **V**$_0$=$\sum_i\sum_j v_{ij}$. Obviously, the summation of the entries in **V** does not equal 1, but the summation of the elements in **W** is equal to 1, i.e., the standard spatial weight matrix satisfies the normalization condition, $\sum_i\sum_j w_{ij}$ =1 (File S1; File S2).

**Table 1 The measures and data sources for empirical analysis of spatial autocorrelation based on Moran's index**

| Measure | Symbol | Meaning | Data source | Year |
|---|---|---|---|---|
| **Distance** | $r_{ij}$ | Interurban distance | Extraction by ArcGIS | 2010 |
| **City size 1** | $x_i^{(1)}$ | Natural logarithm of city population | The fifth and sixth census of China | 2000, 2010 |
| **City size 2** | $x_i^{(2)}$ | Natural logarithm of Nighttime light (NTL) intensity | American NOAA National Centers for Environmental Information (NCEI) | 2000, 2010 |

### 3.2 Empirical results and analysis

The following empirical analysis includes three aspects. First, testify the relationships between the parameters of spatial autocorrelation models and Moran's index. Second, examine the normalized Moran's scatterplots. Third, verify the key formulae for parameter estimation. The third aspect, i.e., verifying representative relations, is just a simple demonstration for readers. The process of spatial autocorrelation modeling and analysis based on Moran's index can be illustrated by a flow chart (Figure 1). The main analytical steps are as follows.

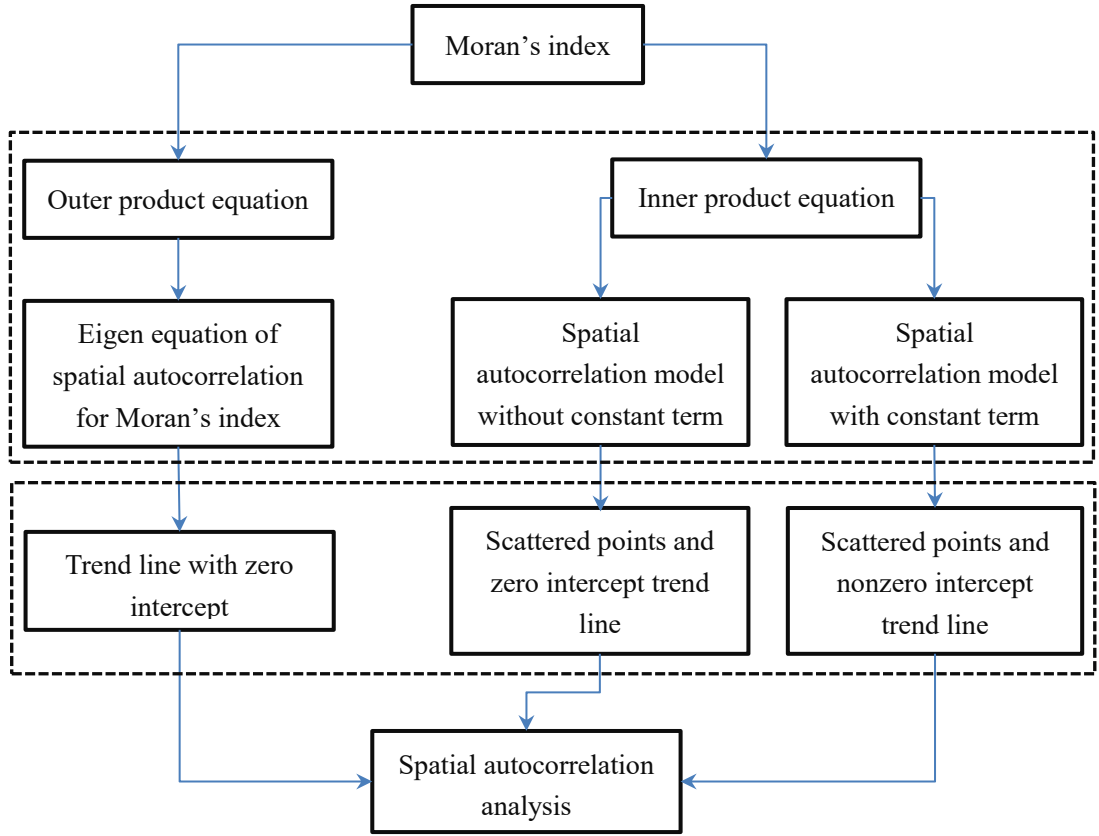

Figure 1 Spatial autocorrelation analysis based on spatial autocorrelation models and normalized Moran's scatterplot: a process of measurement, modeling, scatterplots, and analysis

First of all, the parameter values of spatial autocorrelation models are estimated by using the least squares regression analysis. The logarithms of city population and nighttime light intensity are taken as size measures of cities. The reason for taking logarithm is that the hierarchy of city sizes obeys the law of exponential decay. In other word, Zipf's law can be decomposed into two exponential functions (Chen, 2014). The analysis process consists of five steps as below. **Step 1**: taking logarithm of a city size variable in given time, say, city population in 2010. Thus, we have $y=\ln(x)$, in which $x$ refers to the original variable, and $y$ denotes the logarithmic variable. Standardizing $y$ yields $z$, which is based on the population standard deviation. **Step 2**: normalizing the spatial contiguity matrix to yield spatial weight matrix. As indicated above, the formula is $w_{ij} = v_{ij}/\sum_i\sum_j v_{ij}$, and $v_{ij}=1/r_{ij}$. **Step 3**: estimating Moran's index (slope) and the corresponding constant (intercept). Using equation (6) to make a linear regressive analysis yields the estimated values of the parameters of spatial autocorrelation models. **Step 4**: generating normalized Moran's scatterplot. Adding the standard trend line based on equation (3) and regressive lines based on equations (4) and (5) to Moran's scatterplot based on standardized variable and normalized spatial weight matrix can create improved Moran's scatterplot. **Step 5**: verifying typical formulae and useful relations. The main formulae include the constant term and intercept of spatial autocorrelation models. The basic relation is equation (24), which is useful in spatial autocorrelation modeling.

Now, let's invest spatial autocorrelation models. The independent variable is **z**, and the dependent variable is $n$**Wz**. So, the intercept is $(\mathbf{Wz})^T\mathbf{o}$, and the slope is the Moran's index, $I$. Taking the constant term is account, for the NTL data in 2010, we have a spatial autocorrelation model as follows

$$n(Wz)_i = -0.1427 + 0.1248z_i + e_i, \qquad (37)$$

where $e_i$ denotes residuals. The goodness of fit is $R^2=0.2301$. Equation (37) corresponds to equation (5) and is based on equation (6). According to the results of parameter estimation, Moran's index is about $I=0.1248$, the constant term is -0.1427, which equals the value of $(\mathbf{Wz})^T\mathbf{o}$. This is the mean value of $n\mathbf{Wz}$. Letting the constant term to be zero yields another model as below:

$$n(Wz)_i = 0.1248z_i + e_i^*, \qquad (38)$$

where $e_i^*$ denotes another residuals series. The goodness of fit is $R^2=0.1769$. Equation (38) corresponds to equation (4). This model gives the same value of Moran's index $I=0.1248$. This verifies the following inference: the existence or no of the constant term in the spatial autocorrelation models does not affect the estimation result of Moran index as a slope. In this way, all the model parameter values and corresponding statistics can be worked out (Table 2).

**Table 2 The parameter values and related statistics of two types of spatial autocorrelation models based on Moran's index**

| Measure | Year | Spatial autocorrelation model (I) (Equation (4), without intercept) | | | $R^2$ | Spatial autocorrelation model (II) (Equation (5), with intercept) | | | $R^2$ |
|---|---|---|---|---|---|---|---|---|---|
| | | Parameter | Coefficients | $P$-value | | Parameter | Coefficients | $P$-value | |
| **City population** | 2000 | -- | 0 | -- | 0.0345 | $(\mathbf{Wz})^T\mathbf{o}$ | -0.0839 | 0.0057 | 0.0433 |
| | | $I$ | -0.0347 | 0.2778 | | $I$ | -0.0347 | 0.2303 | |
| | 2010 | -- | 0 | -- | 0.0175 | $(\mathbf{Wz})^T\mathbf{o}$ | -0.0916 | 0.0043 | 0.0223 |
| | | $I$ | -0.0260 | 0.4421 | | $I$ | -0.0260 | 0.3914 | |
| **Nighttime light** | 2000 | -- | 0 | -- | 0.1013 | $(\mathbf{Wz})^T\mathbf{o}$ | -0.1255 | 0.0020 | 0.1311 |
| | | $I$ | 0.0837 | 0.0586 | | $I$ | 0.0837 | 0.0325 | |
| | 2010 | -- | 0 | -- | 0.1769 | $(\mathbf{Wz})^T\mathbf{o}$ | -0.1427 | 0.0011 | 0.2301 |

| | | $I$ | 0.1248 | 0.0106 | | $I$ | 0.1248 | 0.0035 | |

**Note**: The values of spatial autocorrelation indexes come between the intervals determined by the maximum and minimum eigenvalues of spatial weight matrix and the derived matrixes.

Next, let's examine the normalized Moran's scatterplot. Taking the standardized size variable **z** as abscissa axis and *n* times of weighted standardized variable *n***Wz** as ordinate axis, we can generate a Moran's scatterplot. Still take NTL data as an example. Using equation (3) or equation (4), we can add the first trend line to the scatterplot; using equation (5), we can add the second trend line to the scatterplot. The trend line based on equation (3) coincides with the trend line based on equation (4), but they bear different statistical meanings. Equation (3) represents an outer product equation and gives a trend line without intercept. Equation (4) represents an inner product equation and gives a trend line with nonzero intercept (Figure 2). The closer the two trend lines are, the stronger the spatial autocorrelation becomes.

Finally, it is necessary to testify the key formulae derived above. Let's take the NTL intensity data of 2010 as an example. First, three quantities are prepared as follows. The inner product of **Wz** is $(\mathbf{Wz})^T(\mathbf{Wz})=0.0025$, the mean of *n***Wz** is $(\mathbf{Wz})^T\mathbf{o}=-0.1427$. In terms of equations (10) and (23), the inner product of the city sizes and autocorrelation model residuals is

$$\hat{\gamma} = (\mathbf{Wz})^T \mathbf{e} = \sigma_e^2 = 0.0521. \tag{39}$$

According to equation (16), we have

$$\hat{a} = (\mathbf{Wz})^T \mathbf{o} = -0.1427. \tag{40}$$

According to equation (14) or equation (19) or equation (27), we have

$$\hat{a} = \frac{n(\mathbf{Wz})^T \mathbf{Wz} - I^2 - \sigma_e^2}{(\mathbf{Wz})^T \mathbf{o}} = \frac{35 \times 0.0025 - 0.1248^2 - 0.0521}{-0.1427} = -0.1427. \tag{41}$$

According to equation (13) or equation (18) or equation (28), we have

$$\hat{b} = \mathbf{z}^T \mathbf{Wz} = 0.1248. \tag{42}$$

Finally, we can verify equation (24), the results are as below:

$$n(\mathbf{Wz})^T \mathbf{Wz} - I^2 - \sigma_e^2 = 35 \times 0.0025 - 0.1248^2 - 0.0521 = 0.0204. \tag{43}$$

$$((\mathbf{Wz})^T \mathbf{o})^2 = (-0.1427)^2 = 0.0204. \tag{44}$$

This implies that the left side of equation (24) is equal to its right side. Other equations or relations can be verified in the similar way by means of observational data.

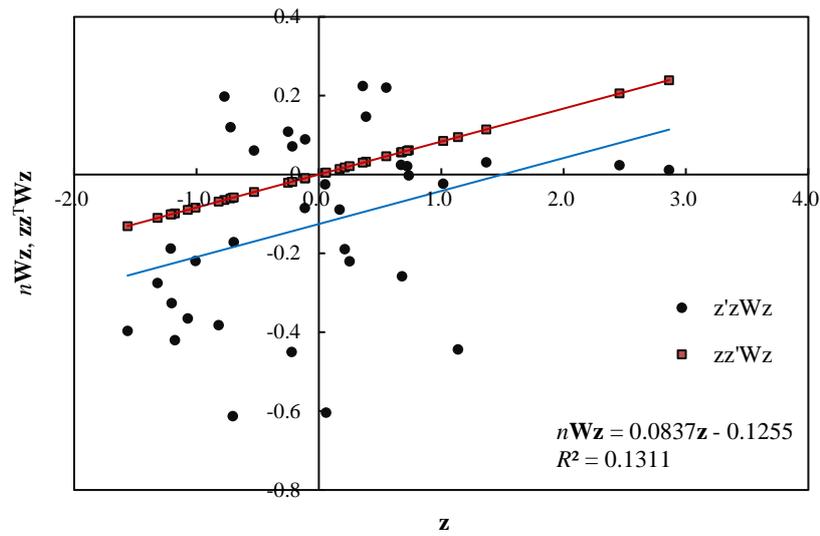

a. 2000

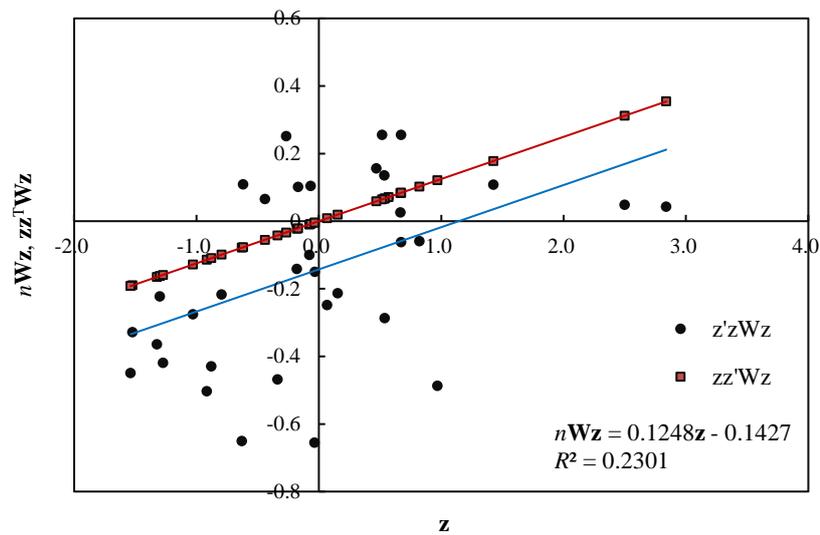

b. 2010

**Figure 2 Normalized Moran's scatterplots for spatial autocorrelation of NTL intensity of cities in Beijing-Tianjin-Hebei region, China**

**Note**: In a normalized Moran's scatterplot, there are two trend lines. One is based on equation (3), and the other is based on equation (5). If the trend line is generated by equation (4), it will coincide with the trend line generated by equation (3). The slope of a trend lines gives the value of Moran's index, while the intercept of the trend line reflect the mean of spatial weighted standardized variables.

Although the objective of this study is not the empirical analysis of spatial autocorrelation of cities in BTH region, we might as well look at the principal data analysis conclusions. First, there is nonlinear spatial autocorrelation rather than linear spatial autocorrelation among these cities. For original variables, which are not taken logarithm, the spatial autocorrelation is not significant for the spurious probability $α=0.05$; in contrast, if the size variables are taken logarithm, there will be significant spatial autocorrelation. This suggests that the spatial autocorrelation is based on allometric relationships rather than linear relationships. Second, for different measurements, spatial autocorrelation effects are different. For urban census population, spatial autocorrelation is not significant; in contrast, for urban NTL data, spatial autocorrelation is significant. This indicates that, despite correlation between city population size and urban NTL intensity, the two measures, population and NTL, cannot be replaced with one another. Third, the intercept value does not influence the slope value, i.e., Moran's *I*. There is a negative correlation relation between the slope values and intercept values. Whether the spatial autocorrelation model retains the constant term or not, the Moran index given by the regression coefficient is not affected. In addition, the model intercept is significant regardless of whether the spatial autocorrelation is significant. Fourth, the spatial autocorrelation evolution trend of urban population sizes and NTL intensity is opposite. From 2000 to 2010, population spatial autocorrelation became weaker, while NTL spatial autocorrelation became stronger. This implies that we can make conventional statistical analysis by means of urban population rather than NTL intensity. After all, the premise of conventional statistical analysis is that there is no significant spatial autocorrelation in a data set. But for NTL datasets, spatial autocorrelation analysis is necessary.

## 4. Discussion

As a spatial autocorrelation coefficient, Moran's index is only a spatial statistic measurement. A measurement is used for description rather than inference. By constructing spatial autocorrelation equations, we can convert the simple measurement into a set of mathematical models (Table 3). Using the models, we can turn the spatial autocorrelation analysis into linear regression analysis which is more familiar to many scientists. In this way, the spatial analysis process become simpler, but more spatial information can be revealed. As shown above, the key models of spatial autocorrelation analysis can be summarized as an inner product equation and an outer product equation. The inner product equations of Moran's index can be regarded as spatial autocorrelation equation, which is actually inverse functions of spatial autoregressive models. By means of the principle of linear algebra, we can derive the expressions of regressive coefficients. The main findings are as follows. First, the intercept, i.e., constant term, is the sum of the elements in the weighted size vector, that is $(\mathbf{Wz})^T\mathbf{o}=\sum(n\mathbf{Wz})/n=\sum\mathbf{Wz}$, and the slope is just equal to Moran's index, that is, $b=I=\mathbf{z}^T\mathbf{Wz}$. Second, removing the intercept does not change the value of the slope. This research leads to new way of understanding Moran's index. First, Moran's index is an eigenvalue of $n$ times of spatial weight matrix in theory, but is limited by the maximum eigenvalue and minimum eigenvalue of the weight matrix in empirical analysis. Second, linear regressive analysis based on least squares calculation can be used to estimate and test Moran's index. Third, the inner product equation can be employed to generate two trend lines in a normalized Moran's scatterplot. The trend line given by the inner product equation without constant term coincides with the trend line given by the characteristic equation based on the outer product. Fourth, combining inner product and outer product equations, we can derive three sets of boundary values for Moran's index. This is helpful to understand the bounds of Moran's index from a new perspective.

Table 3 Similarities and differences between mathematical models and measurements

| Item | Spatial autocorrelation equation | Moran's index |
|---|---|---|
| Attribute | Model | Measurement |
| Objective | Imitation of system structure | Characterization of System Characteristics |
| Mathematical form | Function, equation | Formula |
| Characteristics | There are undetermined parameters, which depend on the algorithm | No undetermined parameters, and the index can be calculated directly |
| Source | Data analysis or theoretical deduction | Definition or construction |
| Main function | Description and inference | Description |

The most important point is that the models developed in this work is useful to understand the principles of spatial autocorrelation and spatial auto-regression from a new angle of view. As indicated above, the inner product equations are actually inverse functions of spatial autoregressive models. A spatial autoregressive model bears analogy with the temporal autoregressive model in the theory of time series analysis. The time lag parameter can be converted into the expression form of time contiguity matrix. However, time is unidirectional and irreversible, and the contiguity matrix based on time lag is asymmetric in principle. Therefore, the vector based on time lag, **Wz**, can only act as independent variable, and the dependent variable must be **z**. In contrast, the spatial process is bidirectional and reversible, and the spatial contiguity matrix is symmetrical in principle -- the asymmetric spatial weight matrix is only a special case. Therefore, the vector based on spatial weight, **Wz**, can not only act as independent variable but also as dependent variable. In this sense, we have both spatial autoregressive models but also spatial autocorrelation models.

The geometric sense of the inner and outer product equations should be made clear. Where the outer product equation is concerned, Moran's index proved to be the only nonzero eigenvalue of **zz$^T$W**. Moran's index is a spatial statistical measure. Under given conditions, the value of the measure is uniquely determinate. The formulae, equation (1), give a certain index value. Equation (3) comes from equation (1). If equation (3) gives more than one nonzero eigenvalue, then Moran's index value will be not unique. This contradicts the nature of Moran index defined in equation (1). Equation (1) gives one nonzero eigenvalue, representing the real extension length of the size vector in one direction. In contrast, where the inner product equation is concerned, we cannot prove the uniqueness of the eigenvalue of $n$**W** based on equation (4). In fact, equation (4) gives $n$ nonzero eigenvalues, representing the maximum extension length of the size vector in $n$ directions.

The characteristic equations of Moran's index can be generalized to other measures for spatial autocorrelation analysis. As we known, many mathematical methods based on quadratic form, symmetric matrix, and reciprocal matrix can be outlined as a pair of equations based on inner product and outer product (Table 4). Based on normalized variables and normalized weight matrixes, Getis-Ord's index can also be expressed as two equations based on inner product and outer product (Chen, 2020). This suggests that the mathematical derivation process for spatial statistic parameters can be generalized to Getis-Ord's index by analogy. Getis-Ord's index reflects spatial autocorrelation from another geographical prospective (Getis and Ord, 1992).

**Table 4 The inner production and outer product equations for a type of mathematical methods**

| Type | Method | Inner product | Outer product |
|---|---|---|---|
| **Spatial statistics** | Moran's index | $\mathbf{z}^T\mathbf{zWz} = n\mathbf{Wz} = I\mathbf{z}$ | $\mathbf{zz}^T\mathbf{Wz} = I\mathbf{z}$ |
| | Getis-Ord's index | $\mathbf{p}^T\mathbf{pWp} = \xi\mathbf{Wp} = G\mathbf{p}$ | $\mathbf{pp}^T\mathbf{Wp} = G\mathbf{p}$ |
| **Multivariable statistics** | Principal component analysis (PCA) and | $(\frac{1}{n}\mathbf{X}^T\mathbf{X})\mathbf{a} = \mathbf{Ra} = \lambda\mathbf{a}$ | $(\frac{1}{n}\mathbf{XX}^T)\mathbf{f} = \lambda\mathbf{f}$ |

| | factor analysis (FA) | | |
|---|---|---|---|
| **System analysis** | Analytic hierarchy process (AHP) | $(\mathbf{WW}^{*T})\mathbf{W} = n\mathbf{W}$ | $\mathbf{W}(\mathbf{W}^{*T}\mathbf{W}) = n\mathbf{W}$ |

**Note**: For Getis-Ord's index $G$, **p** denotes a normalized vector, $\xi=\mathbf{p}^T\mathbf{p}$ refers to the inner product of **p**. For PCA and FA, **X** denotes standardized matrix, **R** is Pearson correlation matrix, $\lambda$ is an eigenvalue, **a** and **f** refer to eigenvectors. For AHP method, **W** denotes weight vector, $\mathbf{W}^*$ is the reciprocal vector of **W**.

Spatial autocorrelation analysis is one of important tools for quantitative analysis of geographical systems. The measures, methods, and theory of spatial autocorrelation has been developed for a long time have made remarkable achievements (Cliff and Ord, 1969; Cliff and Ord, 1973; Cliff and Ord, 1981; Geary, 1954; Getis, 2009; Griffith, 2003; Haining, 1980; Haining, 2009; Sokal and Oden, 1978; Sokal and Thomson, 1987; Tiefelsdorf, 2002). Compared with previous studies on the inner product equation of Moran's index, the novelty of this work is as follows. First, the constant term is taken into account. In previous research, the intercept is ignored. It was proved that intercept is independent of slope. This suggests that whether or not the model has a constant term does not affect the estimation of Moran index. Second, clear calculation formulae of models' parameters were derived. The mathematical structure of both the intercept and slope was brought to light. Third, a new way of understanding the bounds of Moran's index were proposed. Based on spatial weight matrix, the inner product of spatial weight matrix, and the outer product of weighted size vector, we have at least three sets of boundary values of Moran's index. Fourth, normalized Moran's scatterplots were improved. Two trend lines can be added to a Moran's scatterplot based on the inner product equation with intercept and the equation without intercept. Fifth, the most important point is that the models provided a new way of understanding the principles of spatial autocorrelation and spatial auto-regression. The main shortcoming lies in that the spatial autoregressive modeling corresponding to the spatial autocorrelation model has not be discussed. In fact, the spatial autocorrelation models can be treated as inverse spatial autoregressive models. However, due to the limitations of space and topic, the spatial autoregressive modeling based on the inner product equation cannot be deeply explored for the time being.

# 5. Conclusions

The spatial autocorrelation models based on inner product equation and outer product equation of Moran's index is useful in both theoretical development and empirical research of geographical analysis. From the results of mathematical derivation and positive study, the chief conclusions can be drawn as follows. *First, a spatial autocorrelation model is an inner product equation based on Moran's index, and the equation is actually an inverse function of the simplest spatial autoregressive model.* Spatial autocorrelation differs from temporal autocorrelation, so spatial autoregressive process differs from temporal autoregressive process. The difference lies in spatial symmetry and temporal asymmetry. A number of models and methods of spatial analysis were actually developed by analogy with time series analysis. However, time series autocorrelation is a unidirectional process defined in 1-dimensional time, while spatial autocorrelation is a bidirectional process defined in 2-dimensional space. Therefore, a time autoregressive model has no inverse function, while a spatial autoregressive model has an inverse function, that is, spatial autocorrelation model. *Second, inner product equations can be employed to improve normalized Moran's scatterplot.* Based on standardized variable and globally normalized weight matrix, spatial autocorrelation models for Moran's index have a pair of expressions: one is that bears constant term, and the other is that bears no constant term. The existence of constant term does not affect the autocorrelation coefficient of the model. Using the pair of inner product equations, we can improve normalized Moran's scatterplot by adding two trend lines. The trend line based on the zero-intercept autocorrelation model coincides with the standard trend line based on outer product equation, and the trend line based on the autocorrelation model bearing intercept deviates the standard trend line. The slopes of the trend lines give the value of Moran's index, and the distance between the two trend lines reflects the intensity of spatial autocorrelation. *Third, the boundary values of Moran's index depend on the relationships between spatial weight matrix and size vector*. Some scholars believe that the boundary values is -1 and 1, while others believe that it is determined by the maximum and minimum eigenvalues of the spatial weight matrix. However, the actual situation is not so simple. Based on the simplified formula of Moran's index, inner product equation, and outer product equation, we can derive three sets of boundary values. These sets of boundary values are not equivalent to one another. The first one is based on spatial weight matrix, the second one is based on the inner product of the spatial weight matrix, and the third one is based on the outer product of weighted size vector. The actual boundary values must be the intersection of these three sets of boundary values. *Fourth, the regressive analysis based on least squares method can be used to calculate Moran's index and related parameters*. The inner product equations of Moran's index can be treated as spatial autocorrelation models. We have at least four ways of deriving the models' parameters by using the least squares method. The first is solving algebraic equations, the second is regressive coefficient formulae, the third is matrix manipulation, and the fourth is determinant operation based on Cramer's rule. The expressions of parameters given by different methods may be different in form, but the results are equivalent to one another. Considering the internal relationship between autocorrelation models and autoregressive models, we can infer that the least square method can be used to estimate the parameters of spatial autoregressive models.

# Acknowledgements

This research was sponsored by the National Natural Science Foundation of China (Grant No. 42171192). The support is gratefully acknowledged.

# Appendixes

## Appendix 1. The formulae of intercept and slope of linear regressive models

In theory, a univariate linear regression model with a constant term can be expressed as

$$y = a + bx, \tag{A1}$$

where $x$ denotes independent variable, $y$ refers to dependent variable. As for the parameters, $a$ is the intercept, and $b$ is the slope. From equation (A1), we can derive the following relation

$$\overline{y} = a + b\overline{x}, \tag{A2}$$

where $\overline{x}$ and $\overline{y}$ represent he average values of the independent and dependent variables. Suppose that the independent variable is a standardized variable by $z$-score. In this case, the mean of the independent variable is 0. By means of the least squares method, the slope can be calculated by

$$b = \frac{\sum_{i=1}^{n}(x_i - \overline{x})(y_i - \overline{y})}{\sum_{i=1}^{n}(x_i - \overline{x})^2} = \frac{\sum_{i=1}^{n} x_i (y_i - \overline{y})}{\sum_{i=1}^{n} x_i^2} = \frac{\sum_{i=1}^{n} x_i y_i - \overline{y} \sum_{i=1}^{n} x_i}{\sum_{i=1}^{n} x_i^2}. \tag{A3}$$

According to equation (A2), the intercept should be

$$a = \overline{y} - b\overline{x} = \overline{y}. \tag{A4}$$

This suggests that the intercept is equal to the mean of dependent variable. Assuming $x=\mathbf{z}$ and $y=n\mathbf{Wz}$, we can derive the following result: $a=\sum_i(\mathbf{Wz})_i$, $b=I$, where $I$ denotes Moran's index. If the constant term in equation (A1) is zero ($a=0$), the sum of $x_i$ is also equal to zero ($\sum_i x_i = 0$), we have $\bar{y} = a = 0$, then the formula of calculation formula for the slope will change to

$$b = \frac{\sum_{i=1}^{n} x_i y_i}{\sum_{i=1}^{n} x_i^2}. \qquad (A5)$$

Substituting $x=\mathbf{z}$ and $y=n\mathbf{Wz}$ into equation (A5) yields $b=I$.

## Appendix 2. The inner product of residuals of a spatial autocorrelation model

The process of derivation and transformation of the inner product equation of residuals in details is as follows:

$$\begin{aligned}
\mathbf{e}^T\mathbf{e} &= (n\mathbf{Wz} - a\mathbf{o} - b\mathbf{z})^T (n\mathbf{Wz} - a\mathbf{o} - b\mathbf{z}) \\
&= (n(\mathbf{Wz})^T - a\mathbf{o}^T - b\mathbf{z}^T)(n\mathbf{Wz} - a\mathbf{o} - b\mathbf{z}) \\
&= n^2(\mathbf{Wz})^T \mathbf{Wz} - 2an(\mathbf{Wz})^T \mathbf{o} - 2nb(\mathbf{Wz})^T \mathbf{z} + a^2 n + b^2 \mathbf{z}^T \mathbf{z} \\
&= n^2(\mathbf{Wz})^T \mathbf{Wz} - 2an(\mathbf{Wz})^T \mathbf{o} - 2nbI + a^2 n + b^2 n \\
&= n^2(\mathbf{Wz})^T \mathbf{Wz} - 2n((\mathbf{Wz})^T \mathbf{o})^2 - nI^2 + a^2 n \\
&= n^2(\mathbf{Wz})^T \mathbf{Wz} - 2n((\mathbf{Wz})^T \mathbf{o})^2 - nI^2 + n((\mathbf{Wz})^T \mathbf{o})^2 \\
&= n^2(\mathbf{Wz})^T \mathbf{Wz} - n((\mathbf{Wz})^T \mathbf{o})^2 - nI^2 \\
&= n(n(\mathbf{Wz})^T \mathbf{Wz} - ((\mathbf{Wz})^T \mathbf{o})^2 - I^2)
\end{aligned} \qquad (B1)$$

Therefore, we have

$$\frac{1}{n}\mathbf{e}^T\mathbf{e} = n(\mathbf{Wz})^T \mathbf{Wz} - ((\mathbf{Wz})^T \mathbf{o})^2 - I^2, \qquad (B2)$$

which is just the variance of residuals series.

## Supplementary information files

**File S1. Spatial autocorrelation modeling processes and results for BTH cities in 2000.** This file contains the original or preliminarily processed data of urban system the study area of 2000 used in this paper. It provides two complete processes of computing spatial autocorrelation coefficients and gives normalized Moran's scatterplots. (XLSX)

**File S2. Spatial autocorrelation modeling processes and results for BTH cities in 2010**. This file contains the original or preliminarily processed data of urban system the study area of 2010 used in this paper. (XLSX)